\documentclass[11pt,a4paper]{article}

\usepackage{amsmath}
\usepackage{amsfonts}
\usepackage{amssymb}
\usepackage{multicol}
\usepackage{graphicx}
\DeclareGraphicsExtensions{.pdf}

\makeatletter
\newenvironment{figurehere}
  {\def\@captype{figure}}
  {}
\makeatother

\author{Debraj Roy}

\begin{document}
\title{\bf \textsf{The Unruh thermal spectrum through scalar and fermion tunneling}}
\author{
{\bf \textsf{Debraj Roy}}$
$\thanks{E-mail: debraj@bose.res.in}\\
 {\footnotesize \textsf{\it S.N.~Bose National Centre for Basic Sciences,}}\\
 {\footnotesize \textsf{\it Block--JD, Sector III, Salt Lake, Kolkata-700098, India.}}}%
\date{}
\maketitle
\vspace{-1.0 truecm}

\begin{center}\rule{0.9\textwidth}{0.5 pt}\end{center}
\vspace{-0.7 truecm}
\begin{abstract}

The thermal spectrum seen by accelerated observers in Minkowski space vacuum, the Unruh effect, is derived within the tunneling mechanism. This is a new result in this mechanism and it completes the treatment of Unruh effect via tunneling. Both Bose-Einstein and Fermi-Dirac spectrum is derived by considering tunneling of scalar and spin half particles respectively, across the accelerated Rindler horizon. Full solutions of massless Klein-Gordon and Dirac equations in the Rindler metric are employed to achieve this, instead of approximate solutions.\\

\vspace{-0.8 truecm}
\end{abstract} 
\begin{center}\rule{0.9\textwidth}{0.5 pt}\end{center}
\vspace{0.1 truecm}

\begin{multicols}{2}

\section{Introduction}
\label{Intro}

Linearly accelerated observers should detect a thermal background with black-body spectrum in place where an inertial observer detects no particles. This result is known as the Fulling-Davies-Unruh effect \cite{FDU} \footnote{\it\small{for a recent review and extensive references, see \cite{Crispino}}}. Several analysis of non-inertial observers have been done through a quantized field theory construction in the coordinates adapted to an accelerated observer -- Rindler coordinates. Now while analyzing the closely related Hawking effect \cite{Hawking:1974rv} for black-holes, an easier and more conceptually transparent single-particle analysis, the \emph{tunneling mechanism}, was developed \cite{Srinivasan:1998ty, Parikh:1999mf}. This considers a virtual pair of particles formed just inside the horizon, one `ingoing' towards the black-hole i.e. away from horizon, while the other `outgoing' one traveling towards the horizon. Now, as against classical general relativity, this outgoing particle is taken to quantum-mechanically tunnel outside, with a small probability.

The tunneling analysis has also been done in the Unruh effect \cite{Srinivasan:1998ty, Kerner:2006vu, Kerner:2007rr, Akhmedova:2008au} where the Rindler wedges $\mathtt{II}$ and $\mathtt{I}$ (see Fig. \ref{WedgeDia}) act as Black-hole interior and exterior regions, with the accelerated horizon playing the role of the black-hole horizon. However the tunneling mechanism could only derive the Unruh (or Hawking) temperature and not explicitly the spectrum. Recently this drawback was removed \cite{Banerjee:2009wb} through a quantum statistical analysis of a system of particles that are tunneling across the horizon, leading to a clear derivation of the Hawking black-body spectrum, in case of a spherically symmetric black-hole. Other applications of this method may be found in \cite{Umetsu:2009ra} where the Kerr-Newman black hole was considerd, and in \cite{Banerjee:2009sz} where black hole solutions in Lovelock gravity is discussed.

In this paper, I calculate the thermal spectrum in Unruh effect, within the tunneling formalism. The methodology follows  \cite{Banerjee:2009wb}, but with a modification in the calculation of tunneling modes. Instead of the usually applied WKB approximations, I use full solutions of the Klein Gordon and Dirac equations in $(3+1)$-D Rindler spacetime, as the tunneling modes. This is possible as the flat Rindler metric is inherently simpler than, say, the Schwarzschild metric where full solution of the Klein Gordon equation is not known. Thus, this article provides a new and conceptually appealing derivation of the Unruh effect. 


\section{Rindler metric and coordinate extension}
\label{Kruskalization}

The path of constant, linear acceleration (say $\alpha$) in Minkowski spacetime, is described by the hyperbola
\begin{eqnarray}
\label{hyperbola}
X^2 - T^2 =\frac{1}{\alpha^2}
\end{eqnarray}
\vspace{0.3 truecm}
\begin{figurehere}
\centering
\includegraphics[angle=0, width=0.48\textwidth]{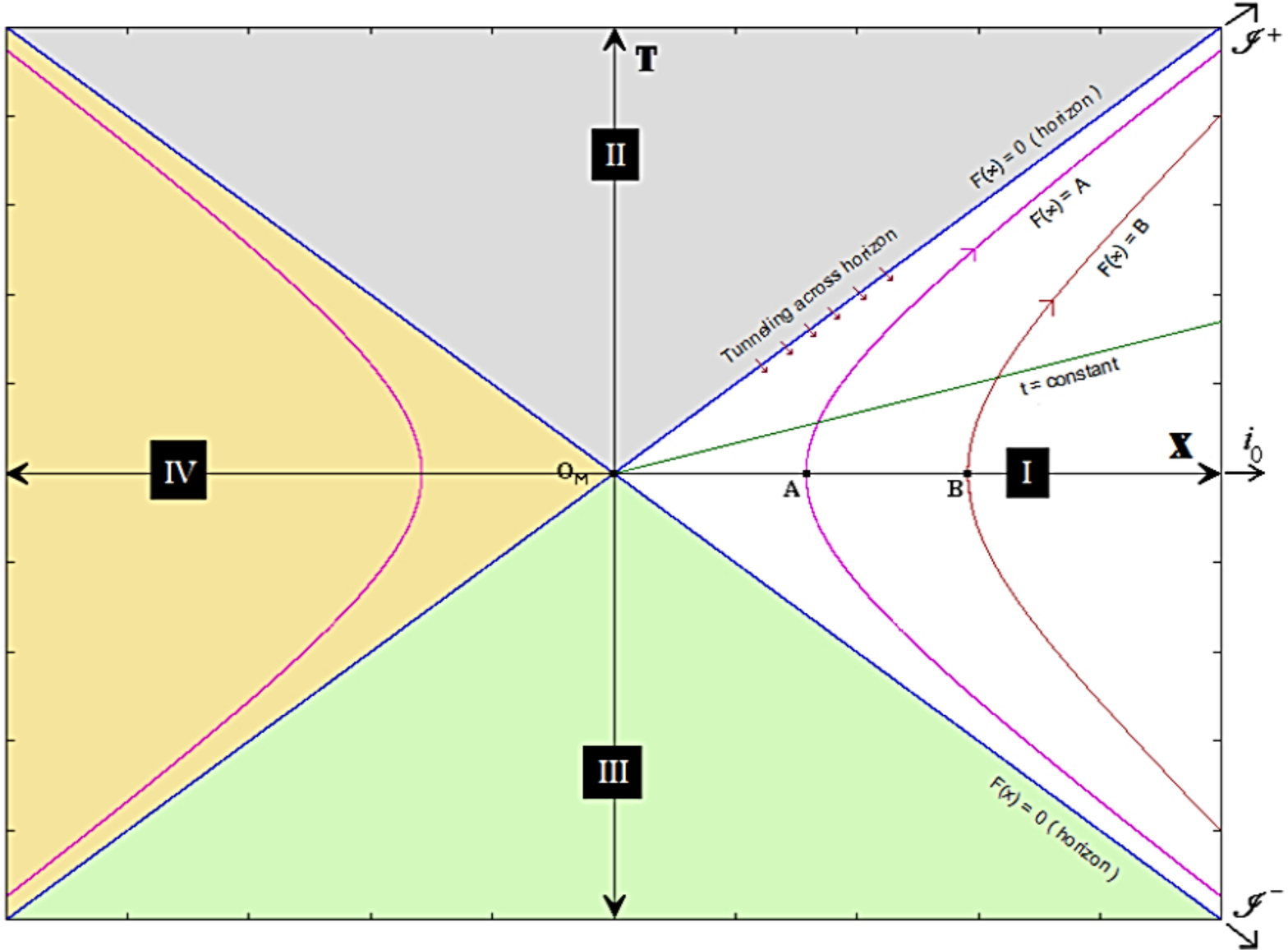}
\caption{\label{WedgeDia}{\it\small\textsf{The Rindler wedges shown on the Minkowski plane. $\textrm(${\bf X},{\bf T}$\textrm)$ and ($x,t$) are Minkowski and Rindler coordinates. The infinities lie outside the diagram, towards the directions shown. Tunneling occurs from wedge $\mathtt{II}$ to $\mathtt{I}$.}}}
\end{figurehere}
\vspace{0.3 truecm}
where $T$ \& $X$ are Minkowski coordinates, with $X$-axis being the direction of acceleration. If  I consider a range ($0<\alpha^2<\infty$) of accelerated observers, the entire `wedges' $\mathtt{I}$ and $\mathtt{IV}$ can be covered with the resultant hyperbolae. These wedges are known as `\emph{Rindler wedges}'. A new coordinate system -- the Rindler coordinates --  can now be set up taking these timelike hyperbolae and corresponding spacelike lines (Fig. \ref{WedgeDia}) as the coordinate lines. They are related to the Minkowski coordinates, in wedges $\mathtt{I}$ and $\mathtt{IV}$ (indicated as subscripts), through the following relations:
\begin{align}
\label{transformations}
\begin{split}
T &=F(x_\mathtt{\scriptscriptstyle I}) \sinh(at_\mathtt{\scriptscriptstyle I})\\
X &=F(x_\mathtt{\scriptscriptstyle I}) \cosh(at_\mathtt{\scriptscriptstyle I})\\
Y &=y_\mathtt{\scriptscriptstyle I}~, \quad Z=z_\mathtt{\scriptscriptstyle I}
\end{split}
\end{align}
where $F(x)^2=1/\alpha^2$ and $a$ is a constant. The Minkowski line element $ds^2=-dT^2+dX^2+dY^2+dZ^2$ can now be written as
\begin{align}
\label{genRind}
ds^2 = -a^2F(x)^2dt^2 + F'(x)^2dx^2 + dy^2 + dz^2.
\end{align}

This is a generalized form of the Rindler metric and for different choices of the function $F(x)$, we can get all the  different forms of Rindler metric seen in literature. Though this metric covers the entire Minkowski plane, the coordinates $t$ and $x$ change in the different wedges. The Rindler horizon occurs at $F(x)^2=0$. The two wedges $\mathtt{I}$ and $\mathtt{II}$ on two sides, act like the black-hole exterior and interior regions respectively, as seen in the case of Schwarzschild solution.

Changing to a tortoise coordinate $x_\star$ appropriate for the Rindler metric (\ref{genRind}) through
\begin{align}
\label{tortoise}
\begin{split}
dx_\star = \frac{F'(x)}{a F(x)} ~dx\\
x_\star = \frac{1}{a} \ln F(x)
\end{split}
\end{align}
the metric (\ref{genRind}) becomes
\begin{align}
\label{GenRinTor}
ds^2 = \left(a~e^{ax\star}\right)^2\left(-dt^2 + dx_\star^2\right) + dy^2 + dz^2.
\end{align}
Returning to equation (\ref{hyperbola}), though the hyperbola
describes a real accelerating particle only in wedge $\mathtt{I}$, 
the hyperbolae can be analytically extended to the wedges $\mathtt{II}$ \&
$\mathtt{III}$ as $X^2 - T^2 =-\frac{1}{\alpha^2}$. The
transformation relating the Rindler coordinate to the Minkowski
coordinate then becomes
\begin{align}
\label{transformationsExt}
\begin{split}
T &=F(x_\mathtt{\scriptscriptstyle II}) \cosh(at_\mathtt{\scriptscriptstyle II})\\
X &=F(x_\mathtt{\scriptscriptstyle II}) \sinh(at_\mathtt{\scriptscriptstyle II})\\
Y &=y_\mathtt{\scriptscriptstyle II}~,  \quad Z=z_\mathtt{\scriptscriptstyle II}.
\end{split}
\end{align}
Now the maximal extension of the Rindler coordinate system $(t,x,\ldots)$ is the Minkowski $(T,X,\ldots)$ which is defined everywhere throughout the four Rindler wedges, i.e. the entire Minkowski plane, irrespective of accelerated horizons which the Rindler observer encounters. The Rindler coordinates however undergo a finite shift through the horizon, as can easily be seen through a comparison of (\ref{transformations}) and (\ref{transformationsExt}). A relation between the $t-x_\star$ coordinate pair outside (Region $\mathtt{I}$) and inside (Region $\mathtt{II}$) can then be written as
\begin{align}
\label{1_2relation}
\begin{split}
t_{\mathtt{\scriptscriptstyle II}}&=t_{\mathtt{\scriptscriptstyle I}}-\frac{i\pi}{2a}\\
x_{\star}{}_{\mathtt{\scriptscriptstyle II}}
&=x_{\star}{}_{\mathtt{\scriptscriptstyle I}}+\frac{i\pi}{2a}.
\end{split}
\end{align}
The $y$ and $z$ coordinates on the other hand, remain unchanged across the horizon. It is to be noted that the transformation $t_{\mathtt{\scriptscriptstyle II}} = t_{\mathtt{\scriptscriptstyle I}}+\frac{i\pi}{2a};\ \ x_{\star}{}_{\mathtt{\scriptscriptstyle II}} = x_{\star}{}_{\mathtt{\scriptscriptstyle I}}-\frac{i\pi}{2a}$ also suffice in relating the coordinates $\mathtt{II}$ and $\mathtt{I}$. However this second pair leads to some problems in taking the classical limit of the tunneling probability, and so is not considered at the accelerated horizon, as will be discussed later. Such transforms were reported earlier in \cite{Akhmedov:2008ru, Banerjee:2008sn}.

\section{Wave function: Scalar particles}
\label{ScalarKG}

The massless Klein Gordon equation $\square \Phi = 0$, written in the Rindler metric (\ref{GenRinTor}), reads
\begin{align}
\label{kg}
\frac{e^{-2 a x_\star}}{a^2}\left[-\partial_t^2\Phi +
\partial_{x_\star}^2\Phi\right] + \partial_y^2\Phi +
\partial_z^2\Phi = 0.
\end{align}
Since the metric is independent of the coordinates $t, y$ \& $z$, I take an ansatz for $\Phi$ as
\begin{align}
\label{sansatz}
\Phi(t,x_\star,y,z)=\phi(x_\star)~e^{-\frac{i}{\hbar}\left(\Omega
t+k_y y+k_z z\right)},
\end{align}
where $\Omega$ is a constant. This $\Omega$ is related to the locally observed energy $\omega$ at some Rindler spacetime point $x_\star$ through a red-shift \cite{Tolman:1934,Carroll:2004st} relation $E_1~V_1=E_2~V_2=\Omega$ connecting the observed energies $E_1$ \& $E_2$ at two different points in a gravitating system at equilibrium. The result ensures that though the observed energies $E$ and the \emph{Tolman red-shift} factor $V$ vary locally as functions of the coordinates (here $x_\star$), their product $\Omega$ is a constant. In a gravitational system in equilibrium, this condition of the constancy of $\Omega$ characterizes the equilibrium, just as is done by temperature in a laboratory thermodynamic system at thermal equilibrium \cite{Tolman:1934}. Here, this locally observed energy $E$ is the energy $\omega$ of the tunneling-particle and the quantity $V$ becomes $\sqrt{|g_{00}|}$. Thus I have
\begin{align}
\label{redshift}
\Omega = \omega\ a e^{a x_\star} = \frac{a\,\omega}{\alpha}
\end{align}
where appropriate definitions $F(x_\star)=1/\alpha=e^{a x_\star}$ (\textit{see metrics} \ref{genRind} \& \ref{GenRinTor}) have been used. Substituting the ansatz (\ref{sansatz}) in the Klein Gordon equation (\ref{kg}), yields the following differential equation in $\phi (x_\star)$
\begin{align}
\label{reducedkg}
\phi''(x_\star)+\left(\frac{\Omega^2}{\hbar^2}-a^2 e^{2 a
x\star}\frac{k_\perp^2}{\hbar^2}\right)\phi(x_\star)=0,
\end{align}
with $k_\perp = \sqrt{k_y^2+k_z^2}$.

Some observations can immediately be made from equation (\ref{reducedkg}). Near the horizon, as $x_\star \rightarrow -\infty$, the term containing $k_\perp$ drops out and a simple harmonic type equation with plane wave solutions is obtained. Again at large spatial distances, $x_\star \rightarrow \infty$, and now the term containing $\Omega^2$ becomes negligible. This leaves an equation with exponentially increasing and decreasing solutions $I_0\left(\frac{k_\perp}{\hbar}e^{a x_\star}\right)$ and $K_0\left(\frac{k_\perp}{\hbar}e^{a x_\star}\right)$, where $I_0$ and $K_0$ represent the zero-th order modified Bessel functions of first and second types respectively. Thus throwing away the $I_0$ solutions, we have an exponentially vanishing solution at infinity, in $K_0$. Similar conclusions have also been reached at by Boulware \cite{Boulware:1974dm}.

The solution of the full equation (\ref{reducedkg}) that is well defined through the horizon is,
\begin{align}
\label{phix}
\phi(x_\star)=&A_-\ e^{\frac{\pi\Omega}{2a\hbar}}\ \Gamma\left(1-\frac{i \Omega}
{a\hbar}\right) I_{-\frac{i \Omega}{a\hbar}}\left(\frac{k_\perp}{\hbar}
e^{a x_\star}\right)\ +\nonumber\\
&A_+\ e^{-\frac{\pi\Omega}{2a\hbar}}\ \Gamma\left(1+\frac{i \Omega}
{a\hbar}\right) I_{\frac{i \Omega}{a\hbar}}\left(\frac{k_\perp}
{\hbar}e^{a x_\star}\right)
\end{align}
where $A_\mp$ are arbitrary integration constants. For small arguments, the appropriate expansion of the modified Bessel function is $I_\nu(z)\simeq\frac{\left(z/2\right)^\nu} {\Gamma(1+\nu)}$. This holds if $k_\perp\ll \Omega$ and also especially near the horizon where $x_\star\rightarrow-\infty$. Therefore (\ref{phix}) simplifies to
\begin{align*}
\phi(x_\star)\simeq A_\mp\ e^{\pm\frac{\pi\Omega}{2a\hbar}}
\left(\frac{k_\perp}{2\hbar}\right)^{\mp\frac{i\Omega}{a\hbar}}\
e^{\mp\frac{i}{\hbar}\Omega x_\star}.
\end{align*}
The total wave function $\Phi(t,x_\star,y,z)$ near the horizon is then
\begin{align}
\label{totswave}
\Phi(t,x_\star,y,z)=&B_{\mathsf{\scriptscriptstyle in}}e^{-\frac{i}{\hbar}
\left[\Omega(t+x_\star) + k_y y+k_z z \right]}\ +\nonumber\\
&B_{\mathsf{\scriptscriptstyle out}} e^{-\frac{i}{\hbar}\left[\Omega(t-x_\star)
+ k_y y+k_z z \right]},
\end{align}
with all the constants clubbed together within $B_{\mathsf{in/out}}$. The subscript `${\mathsf{in}}$' here stands for the ingoing mode which travels toward the accelerated horizon at $x_\star=-\infty$, while the subscript `${\mathsf{out}}$' stands for the outgoing mode traveling away from horizon, i.e. towards $x_\star=\infty$.


\section{Wave function: Spin $\frac{1}{2}$ particles}
\label{FermionDirac}

Spinors are introduced on a general curved spacetime with metric $g$, by going to a local Lorentz frame with metric $\eta$, at each spacetime point \cite{Birrell:1982ix}. This procedure is best done by constructing tetrad fields which map curved spacetime tensors to local Lorentz frame and vice-versa. For a Rindler metric in the tortoise-like coordinate system  (\ref{GenRinTor}), the tetrad field $V^a_\mu$ is defined through the relation
\begin{align*}
g_{\mu\nu}=V^a_\mu\ V^b_\nu\ \eta_{ab}
\end{align*}
where latin ($a,b,\ldots$) and greek ($\mu,\nu,\ldots$) letters run over local Lorentz and curved space indices respectively. The explicit choice of the tetrad field $V^a_\mu$ adopted here is
\begin{align}
\label{tetrad}
V_\mu^a={\rm diag}(a e^{a x_\star}, a e^{a x_\star}, 1, 1)
\end{align}
and the metric signature, both global and local, is kept same as ($-,+,+,+$).

The massless Dirac equation is written as \cite{Giammatteo, Crispino}
\begin{equation}
\label{DiracEq}
\left[\gamma^a \ V_a^\mu\ (\partial _\mu+\Gamma_\mu)\right]\Psi=0
\end{equation}
where $\gamma^a $ are the Dirac matrices obeying the usual algebra
$\left[\gamma^a,\gamma^b\right]=2\eta^{ab}$ and $\Gamma_\mu$ are
connection coefficients given by
\begin{align*}
\begin{split}
\Gamma_\mu &= \frac{1}{2}\ \Sigma^{ab}\ V_a^\nu\ V_{b\nu;\mu}\\
\Sigma^{ab}&=\frac{1}{4}\left[\gamma^a,\gamma^b\right].
\end{split}
\end{align*}
The covariant derivative over the curved space index of the tetrad is defined in the usual way $V_{b\nu;\mu} = \partial_{\mu}V_{b\nu} - \Gamma^{\alpha}_{\mu\nu} V_{b\alpha}$, where $\Gamma^{\alpha}_{\mu\nu}$ is the Christoffel symbol. On using the properties of $\gamma$ matrices and the diagonal choice of the tetrad (\ref{tetrad}), the spin-connection becomes $\Gamma_\mu = -\frac{1}{2}\ \Sigma^{ab}\ \Gamma^\lambda_{\mu\nu}\ V^\nu_a\ V_{b \lambda}$. The Dirac equation (\ref{DiracEq}) then turns out to be
\begin{align}
\label{DiracEq2}
\big[~\left(\partial_t-a\Sigma^{01}\right)-\gamma^0\gamma^1 &\partial_{x_\star}\nonumber\\
- ae^{ax_\star}\big(\gamma^0\gamma^2 &\partial_y +
\gamma^0\gamma^3\partial_z \big)~\big]~\Psi=0.
\end{align}
The ansatz for the spinor $\Psi$ is taken as
\begin{align}
\label{DansatzPsi}
\Psi(t,x_\star,y,z)&=\psi(x_\star)e^{-\frac{i}{\hbar}
 \left(\Omega t+k_y y+k_z z\right)}\nonumber\\
\psi(x_\star)&=\left[\begin{matrix}A(x_\star)\\0\\B(x_\star)\\0\end{matrix}\right].
\end{align}
Upon using this ansatz, equation (\ref{DiracEq2}) can be cast into a Schr\"{o}dinger like equation
\begin{align}
\label{Schrod}
\hat{H}\ \psi(x_\star)=\Omega\ \psi(x_\star)
\end{align}
where the Hamiltonian-like operator $\hat{H}$ is
\begin{align}
\label{hamilt}
\hat{H}=i\hbar\big(a\Sigma^{01}&+\gamma^0\gamma^1\partial_{x_\star}\big)\nonumber\\
&+ae^{a x_\star}\big(k_y\gamma^0\gamma^2+k_z\gamma^0\gamma^3\big).
\end{align}
The above equation, on squaring, gives $\hat{H}^2\psi=\Omega^2\psi$. Now using the ansatz (\ref{DansatzPsi}) and adopting the convention for the gamma matrices as $\gamma^0=\left(\begin{smallmatrix}-i&0\\0&i\end{smallmatrix}\right)$, $\gamma^j=\left(\begin{smallmatrix}0&-i\sigma^j\\i\sigma^j&0\end{smallmatrix}\right)$ (where $j=1,2,3$ and $\sigma^j$ are the Pauli matrices), the following equation for the spinor component functions is obtained
\begin{align}
\label{ABEqns}
\blacklozenge''+a\,\blacklozenge'+\Biggl[\frac{\Omega^2}{\hbar^2}-a^2\,e^{2ax_\star}\frac{k_\perp^2}{\hbar^2}+\frac{a^2}{4} \Biggl]\blacklozenge=0.
\end{align}
Here $\blacklozenge$ stands for the functions $A(x_\star)$ and $B(x_\star)$. As in the case of scalar particles (\ref{reducedkg}), a study of asymptotic behaviour of this equation show that solutions near the horizon are oscillatory, and that near to infinity are vanishing in nature. Solution for the full equation (\ref{ABEqns}) turns out to be
\begin{align}
\label{ABSolns}
\blacklozenge={}&e^{-\frac{ax_\star}{2}}\biggl[M_-^{(\blacklozenge)}
 \ e^{\frac{\pi\Omega}{2a\hbar}}\ \Gamma\left(1-\frac{i \Omega}{a\hbar}\right)
  I_{-\frac{i \Omega}{a\hbar}}\left(\frac{k_\perp}{\hbar}e^{a x_\star}\right)\ \nonumber\\
+{}&M_+^{(\blacklozenge)}\ e^{-\frac{\pi\Omega}{2a\hbar}}\ \Gamma
\left(1+\frac{i \Omega}{a\hbar}\right) I_{\frac{i
\Omega}{a\hbar}}\left(\frac{k_\perp}{\hbar}e^{a
x_\star}\right)\biggl].
\end{align}
It is to be noted that both (\ref{ABSolns}) and (\ref{phix}) are full solutions of the respective differential equations (\ref{ABEqns}) and (\ref{reducedkg}), without any approximations of parameters. This can be easily verified by substituting back these solutions in the original differential equations, to see that they satisfy them without using any approximations on the parameters. However, an approximation is used to get a simpler solution from (\ref{ABSolns}), by using the appropriate expansion of the modified Bessel function for small arguments which is given as $I_\nu(z)\simeq\frac{\left(z/2\right)^\nu} {\Gamma(1+\nu)}$. This holds if $k_\perp\ll \Omega$ and also especially near the horizon where $x_\star\rightarrow-\infty$. So, for the region near to the horizon, the total spinor $\Psi$ can finally be written as
\begin{align}
\label{DiracFinSol}
\Psi(t,x_\star,y,z)={}&\xi_{\mathsf{\scriptscriptstyle in}}e^{-\frac{ax_\star}{2}}
e^{-\frac{i}{\hbar}\left[\Omega(t+x_\star) + k_y y+k_z z \right]}\ +\nonumber\\
&\xi_{\mathsf{\scriptscriptstyle out}}e^{-\frac{ax_\star}{2}} e^{-\frac{i}{\hbar}
\left[\Omega(t-x_\star) + k_y y+k_z z \right]},
\end{align}
with $\xi_{\mathsf{\scriptscriptstyle in/out}}$ being constant spinors and the subscript `$\mathsf{in}$' or `$\mathsf{out}$' standing for `ingoing' or `outgoing' modes.

\section{Unruh effect through tunneling: thermal spectrum and temperature}
\label{spectrum}

Uptil now, I had found single-particle wave-functions for bosons and fermions, by solving for the Klein-Gordon and Dirac equations in the Rindler coordinates. These solutions are valid in both Rindler wedges $\mathtt{I}$ and $\mathtt{II}$, but in coordinates $(t_{\mathtt{\scriptscriptstyle I}}, x_{\mathtt{\scriptscriptstyle I}})$ and $(t_{\mathtt{\scriptscriptstyle II}}, x_{\mathtt{\scriptscriptstyle II}})$ respectively. Classically, both the ingoing and outgoing modes (say of a virtual pair instantaneously produced) in wedge $\mathtt{II}$ are trapped, as nothing from inside can cross the horizon and come out to wedge $\mathtt{I}$. However in the tunneling mechanism, an outgoing particle can quantum-mechanically tunnel out across the horizon and into wedge $\mathtt{I}$. This process occurs with a probability given by the Maxwell term $e^{-\frac{2 \pi \omega}{\hbar a}}$, that appropriately goes to zero in the classical ($\hbar \rightarrow 0$) limit. Now to find the energy distribution of a collection of such particles, I will (following \cite{Banerjee:2009wb}) construct a suitable density matrix for both bosons and fermions, and find out the average number of particles having some particular energy $\omega$.

Starting first with bosonic particles, the relation between inside and outside wave-functions is found by using the connection between coordinates (\ref{1_2relation}) in equation (\ref{totswave}) for the modes.
\begin{align}
\label{scalarOutIn}
B_{\mathsf{\scriptscriptstyle in}}e^{-\frac{i}{\hbar}\left[\Omega(t_{\mathtt{\scriptscriptstyle
II}}+x_\star{}_{\mathtt{\scriptscriptstyle II}}) + k_y
y_{\mathtt{\scriptscriptstyle II}}+ k_z
z_{\mathtt{\scriptscriptstyle II}}\right]}{}& = \nonumber\\
B_{\mathsf{\scriptscriptstyle in}}{}&e^{-\frac{i}{\hbar}
\left[\Omega(t_{\mathtt{\scriptscriptstyle
I}}+x_\star{}_{\mathtt{\scriptscriptstyle I}}) + k_y
y_{\mathtt{\scriptscriptstyle I}}+
k_z z_{\mathtt{\scriptscriptstyle I}}\right]} \nonumber\\%
B_{\mathsf{\scriptscriptstyle out}}e^{-\frac{i}{\hbar}
\left[\Omega(t_{\mathtt{\scriptscriptstyle
II}}-x_\star{}_{\mathtt{\scriptscriptstyle II}}) + k_y
y_{\mathtt{\scriptscriptstyle II}}+ k_z
z_{\mathtt{\scriptscriptstyle II}} \right]}{}& =\nonumber\\
\left(e^{-\frac{\pi\Omega}{a\hbar}}\right) B_{\mathsf{\scriptscriptstyle out}}{}&e^{-\frac{i}{\hbar}\left[
\Omega(t_{\mathtt{\scriptscriptstyle
I}}-x_\star{}_{\mathtt{\scriptscriptstyle I}}) + k_y
y_{\mathtt{\scriptscriptstyle I}}+ k_z
z_{\mathtt{\scriptscriptstyle I}} \right]}
\end{align}

Now, let there be `$n$' pair of free particles (ingoing and outgoing) in wedge $\mathtt{II}$. The total state of this system of particles, with each being described by the sector $\mathtt{II}$ modes in equation (\ref{totswave}), is
\begin{align}
\label{scalTotn1}
|\chi_{\scriptscriptstyle B}\rangle &= N_B \displaystyle\sum_{n=0}^\infty |n_\mathtt{\scriptscriptstyle II}^\mathsf{\scriptscriptstyle in}\rangle \otimes |n_\mathtt{\scriptscriptstyle II}^\mathsf{\scriptscriptstyle out}\rangle\nonumber\\
&= N_B \displaystyle\sum_{n=0}^\infty \left(e^{-\frac{n\pi\Omega}{a\hbar}} \right)|n_\mathtt{\scriptscriptstyle I}^\mathsf{\scriptscriptstyle in}\rangle \otimes |n_\mathtt{\scriptscriptstyle I}^\mathsf{\scriptscriptstyle out}\rangle
\end{align}
where $N_B$ is a normalization constant defined through $\langle\chi_{\scriptscriptstyle B}|\chi_{\scriptscriptstyle B}\rangle=1$. The sum over $n$ runs from $0$ to $\infty$ here, in the case of bosons. But in case of fermions, as will be used later, $n$ is limited to $0$ and $1$ by Pauli's exclusion principle. The normalization of $|\chi_{\scriptscriptstyle B}\rangle$ leads to
\begin{align}
\label{scalNfind}
N_B^2\displaystyle\sum_{n,m=0}^\infty &e^{-\frac{(n+m)\pi\Omega}{a\hbar}} \Big(\langle m_\mathtt{\scriptscriptstyle I}^\mathsf{\scriptscriptstyle out}| \otimes \langle m_\mathtt{\scriptscriptstyle I}^\mathsf{\scriptscriptstyle in}|\Big) \Big(| n_\mathtt{\scriptscriptstyle I}^\mathsf{\scriptscriptstyle in}\rangle \otimes |n_\mathtt{\scriptscriptstyle I}^\mathsf{\scriptscriptstyle out}\rangle\Big)=1\nonumber\\
\Rightarrow N_B^2&=\left[\displaystyle\sum_{n=0}^\infty e^{-\frac{2\pi n \Omega}{a\hbar}}\right]^{-1},
\end{align}
and finally for bosons, we have
\begin{align}
\label{scalN}
N_B=\left( 1 - e^{-\frac{2\pi\Omega}{a\hbar}} \right)^{\frac{1}{2}}.
\end{align}

The density matrix operator for this system of bosons is defined as usual
\begin{align}
\label{scalDens}
\hat{\rho}_{\scriptscriptstyle B}=\Big(1-e^{-\frac{2\pi\Omega}{a\hbar}}\Big)
 \displaystyle\sum_{n,m=0}^\infty e^{-\frac{(n+m)\pi\Omega}{a\hbar}} |n_\mathtt{\scriptscriptstyle I}^\mathsf{\scriptscriptstyle in}\rangle &\otimes |n_\mathtt{\scriptscriptstyle I}^\mathsf{\scriptscriptstyle out}\rangle\nonumber\\
\langle m_\mathtt{\scriptscriptstyle I}^\mathsf{\scriptscriptstyle out}|
&\otimes\langle m_\mathtt{\scriptscriptstyle I}^\mathsf{\scriptscriptstyle in}|.
\end{align}
Since ingoing waves are trapped within the horizon and outgoing particles contribute to spectrum, we trace out over the ingoing particles, to form the density matrix for outgoing modes,
\begin{align}
\label{scalDensOut}
\hat{\rho}_{\scriptscriptstyle B}^\mathsf{\scriptscriptstyle out}=\Big(1-e^{-\frac{2\pi\Omega}{a\hbar}}\Big)\displaystyle\sum_{n=0}^\infty e^{-\frac{2\pi n\Omega}{a\hbar}} |n_\mathtt{\scriptscriptstyle I}^\mathsf{\scriptscriptstyle out}\rangle \langle n_\mathtt{\scriptscriptstyle I}^\mathsf{\scriptscriptstyle out}|.
\end{align}
The average number of outgoing particles is then calculated as
\begin{align}
\label{BoseSpect}
\langle\hat{n}_{\scriptscriptstyle B}\rangle &= \text{Tr}^\mathsf{\scriptscriptstyle out}\left[\hat{n}_{\scriptscriptstyle B}\hat{\rho}_{\scriptscriptstyle B}^\mathsf{\scriptscriptstyle out}\right]=\Big(1-e^{-\frac{2\pi\Omega}{a\hbar}}\Big)\displaystyle\sum_{n=0}^\infty n e^{-\frac{2\pi n\Omega}{a\hbar}}\nonumber\\
&=\frac{1}{e^{\frac{2\pi\Omega}{a \hbar}}-1}\nonumber\\
&=\frac{1}{e^{\frac{2\pi\omega}{\hbar \alpha}}-1}
\end{align}
where in the last step, the red-shift definition of $\Omega$ given in (\ref{redshift}) was used. This is immediately recognizable as the Bose-Einstein distribution for a black body at a temperature $T_{\scriptscriptstyle U}$, the Unruh temperature \cite{FDU}, given as
\begin{align}
\label{UnruhTemp}
T_{\scriptscriptstyle U}=\frac{\hbar \alpha}{2 \pi}
\end{align}
with $\alpha$ being the local acceleration.

For fermions, the same method goes through step by step. The connection between spinorial wave-functions in wedges $\mathtt{II}$ and $\mathtt{I}$ is obtained by using (\ref{1_2relation}) in (\ref{DiracFinSol}).
\begin{align}
\label{fermOutIn}
\xi_{\mathsf{\scriptscriptstyle in}}e^{-\frac{i}{\hbar}\left[\Omega(t_{\mathtt{\scriptscriptstyle
II}}+x_\star{}_{\mathtt{\scriptscriptstyle II}}) + k_y
y_{\mathtt{\scriptscriptstyle II}}+ k_z
z_{\mathtt{\scriptscriptstyle II}}\right]}{}& = \nonumber\\
\xi_{\mathsf{\scriptscriptstyle in}}{}&e^{-\frac{i}{\hbar}
\left[\Omega(t_{\mathtt{\scriptscriptstyle
I}}+x_\star{}_{\mathtt{\scriptscriptstyle I}}) + k_y
y_{\mathtt{\scriptscriptstyle I}}+
k_z z_{\mathtt{\scriptscriptstyle I}}\right]} \nonumber\\%
\xi_{\mathsf{\scriptscriptstyle out}}e^{-\frac{i}{\hbar}
\left[\Omega(t_{\mathtt{\scriptscriptstyle
II}}-x_\star{}_{\mathtt{\scriptscriptstyle II}}) + k_y
y_{\mathtt{\scriptscriptstyle II}}+ k_z
z_{\mathtt{\scriptscriptstyle II}} \right]}{}& =\nonumber\\
\left(e^{-\frac{\pi\Omega}{a\hbar}}\right) \xi_{\mathsf{\scriptscriptstyle out}}{}&e^{-\frac{i}{\hbar}\left[
\Omega(t_{\mathtt{\scriptscriptstyle
I}}-x_\star{}_{\mathtt{\scriptscriptstyle I}}) + k_y
y_{\mathtt{\scriptscriptstyle I}}+ k_z
z_{\mathtt{\scriptscriptstyle I}} \right]}
\end{align}
The normalization of the total state ket for fermions
\begin{align}
\label{fermTotn1}
|\chi_{\scriptscriptstyle F}\rangle &= N_F \displaystyle\sum_{n=0}^1 |n_\mathtt{\scriptscriptstyle II}^\mathsf{\scriptscriptstyle in}\rangle \otimes |n_\mathtt{\scriptscriptstyle II}^\mathsf{\scriptscriptstyle out}\rangle\nonumber\\
&= N_F \displaystyle\sum_{n=0}^1 \left(e^{-\frac{n\pi\Omega}{a\hbar}} \right)|n_\mathtt{\scriptscriptstyle I}^\mathsf{\scriptscriptstyle in}\rangle \otimes |n_\mathtt{\scriptscriptstyle I}^\mathsf{\scriptscriptstyle out}\rangle
\end{align}
is again done through $\langle\chi_{\scriptscriptstyle F}|\chi_{\scriptscriptstyle F}\rangle=1$. The sum over number of particles in a given state, in fermionic calculations, always run from $0$ to $1$, following Pauli's exclusion principle. The normalization constant $N_F$ turns out to be
\begin{align}
\label{fermN}
N_F=\frac{1}{\sqrt{1+e^{-\frac{2\pi \Omega}{a\hbar}}}}.
\end{align}
The density operator for fermions $\hat{\rho}_{\scriptscriptstyle F}$ is defined as $|\chi_{\scriptscriptstyle F}\rangle  \langle\chi_{\scriptscriptstyle F}|$. Using the outgoing fermionic density operator $\hat{\rho}_{\scriptscriptstyle F}^\mathsf{\scriptscriptstyle out}$ the spectrum is calculated as average number of outgoing particles
\begin{align}
\label{Fermspect}
\langle\hat{n}_{\scriptscriptstyle F}\rangle &= \text{Tr}^\mathsf{\scriptscriptstyle out}\left[\hat{n}_{\scriptscriptstyle F}\hat{\rho}_{\scriptscriptstyle F}^\mathsf{\scriptscriptstyle out}\right]=\frac{1}{1+e^{-\frac{2\pi \Omega}{a\hbar}}}\displaystyle\sum_{n=0}^1 n e^{-\frac{2\pi n\Omega}{a\hbar}}\nonumber\\
&=\frac{1}{e^{\frac{2\pi\Omega}{ah}}+1}\nonumber\\
&=\frac{1}{e^{\frac{2\pi\omega}{\hbar \alpha}}+1}
\end{align}
where equation (\ref{redshift}) was used in the last step. This is the Fermi-Dirac distribution, at the Unruh temperature $T_{\scriptscriptstyle U}$ defined in (\ref{UnruhTemp}), and constitutes the Unruh effect for accelerated fermions \cite{Candelas}.


\section{Discussions}
\label{discuss}

In this paper, I calculated the thermal spectrum for Unruh effect within the tunneling mechanism. The Unruh temperature was identified via a comparison between the calculated thermal distribution and the standard form of Bose or Fermi distributions. 

However, the temperature can also be calculated directly via tunneling. Say, when an observer in Rindler wedge $\mathtt{I}$ observes an outgoing particle, coming from within wedge $\mathtt{II}$, he will see the wave function from inside wedge $\mathtt{II}$ with a factor as shown in equation (\ref{scalarOutIn}) for scalar particles, and in (\ref{fermOutIn}) for fermions. The ingoing wave, however, does not change by such a factor between the two wedges. To calculate the temperature, we can now use the principle of detailed balance, $\frac{P_{\text{out}}}{P_{\text{in}}}=e^{-\frac{\omega}{T}}$, where $P_{\text{out/in}}=|\text{wave-function}|^2$ is the outgoing/ingoing probability, $\omega$ is the observed energy at that point and $T$ is the temperature. This gives the Unruh temperature $T_U=\frac{\hbar\alpha}{2\pi}$ after suitably using the red-shift definition of $\Omega$ given in (\ref{redshift}) at the observer's point. For a more detailed discussion on the connection between the earlier approaches to tunneling and the one used in this paper, the reader is directed to \cite{Banerjee:2009wb, Banerjee:2008sn}.

Another point to note concerns an apparent ambiguity in the sign of the factors chosen to connect the coordinates between wedges $\mathtt{I}$ and $\mathtt{II}$ in equation (\ref{1_2relation}). It can be verified that the relations $t_{\mathtt{\scriptscriptstyle II}} = t_{\mathtt{\scriptscriptstyle I}}+\frac{i\pi}{2a};\ \ x_{\star}{}_{\mathtt{\scriptscriptstyle II}} = x_{\star}{}_{\mathtt{\scriptscriptstyle I}}-\frac{i\pi}{2a}$, which have a change in sign between the $t$ and $x_\star$ term signs, also connect the transformations (\ref{transformations}) and (\ref{transformationsExt}). However this set is not employed as it produces an exponential factor with a wrong exponent sign $\Big(e^{\tfrac{\pi\Omega}{a\hbar}}\Big)$ in the relations between wave-functions in wedges $\mathtt{I}$ and $\mathtt{II}$ (\ref{scalarOutIn} \& \ref{fermOutIn}). With this alternate sign, the probability of outgoing particles to tunnel out across the horizon, diverges at the classical limit of $\hbar\rightarrow 0$. This is clearly unacceptable. So this obvious criterion of conformity with classical limit is used to remove the above said ambiguity of signs.

\paragraph*{Acknowledgments:} I thank Prof. R. Banerjee for suggesting the problem and for constant encouragement throughout. I also thank Mr. B.R. Majhi and Mr. S. Kulkarni for discussions.


\end{multicols}

\end{document}